\documentclass[twocolumn,apl,epsf,superscriptaddress]{revtex4}
\usepackage{amsmath}
\usepackage{amsfonts}
\usepackage{epsfig}
\usepackage{epsf}
\usepackage{array}
\usepackage{color}
\usepackage{ulem}

\begin{document}

\title{Localized-itinerant dichotomy and unconventional magnetism in SrRu$_2$O$_6$}
\altaffiliation{
Copyright  notice: This  manuscript  has  been  authored  by  UT-Battelle, LLC under Contract No. DE-AC05-00OR22725 with the U.S.  Department  of  Energy.   
The  United  States  Government  retains  and  the  publisher,  by  accepting  the  article  for  publication, 
acknowledges  that  the  United  States  Government  retains  a  non-exclusive, paid-up, irrevocable, world-wide license to publish or reproduce the published form of this manuscript, 
or allow others to do so, for United States Government purposes.  
The Department of Energy will provide public access to these results of federally sponsored  research  in  accordance  with  the  DOE  Public  Access  Plan 
(http://energy.gov/downloads/doe-public-access-plan)}

\author{Satoshi Okamoto}
\altaffiliation{okapon@ornl.gov}
\affiliation{Materials Science and Technology Division, Oak Ridge National Laboratory, Oak Ridge, Tennessee 37831, USA}
\author{Masayuki Ochi}
\affiliation{Department of Physics, Graduate School of Science, Osaka University, Osaka 560-0043, Japan}
\affiliation{RIKEN Center for Emergent Matter Science (CEMS), Hirosawa, Wako, Saitama 351-0198, Japan}
\author{Ryotaro Arita}
\affiliation{RIKEN Center for Emergent Matter Science (CEMS), Hirosawa, Wako, Saitama 351-0198, Japan}
\author{Jiaqiang Yan}
\affiliation{Materials Science and Technology Division, Oak Ridge National Laboratory, Oak Ridge, Tennessee 37831, USA}
\author{Nandini Trivedi}
\affiliation{Department of Physics, The Ohio State University, Columbus, Ohio 43210, USA}

\begin{abstract}
Electron correlations tend to generate local magnetic moments that usually order if the lattices are not too frustrated. 
The hexagonal compound SrRu$_2$O$_6$ has a relatively high Neel temperature but small local moments, 
which seem to be at odds with the nominal valence of Ru$^{5+}$ in the $t_{2g}^3$ configuration. 
Here, we investigate the electronic property of SrRu$_2$O$_6$ using density functional theory (DFT) combined with dynamical-mean-field theory (DMFT). 
We find that the strong hybridization between Ru $d$ and O $p$ states results in a Ru valence that is closer to $+4$,
leading to the small ordered moment $\sim1.2\mu_B$. 
While this is consistent with a DFT prediction, correlation effects are found to play a significant role. 
The local moment per Ru site remains finite $\sim2.3\mu_B$ in the whole temperature range investigated. 
Due to the lower symmetry, the $t_{2g}$ manifold is split and the quasiparticle weight is renormalized significantly in the $a_{1g}$ state, 
while the renormalization in $e_g'$ states is about a factor of 2--3 weaker. 
Our theoretical Neel temperature $\sim700$~K is in reasonable agreement with experimental observations. 
SrRu$_2$O$_6$ is a unique system in which localized and itinerant electrons coexist with the proximity to an orbitally-selective Mott transition within the $t_{2g}$ sector. 
\end{abstract}


\maketitle

\date{\today }


\section*{Introduction}

For systems with an odd number of electrons per unit cell, correlation effects can lead to insulating magnetic ground states. 
Depending on the relative strength between the interactions $U$ and the electron bandwidth or kinetic energy $W$, 
there are two classes of insulators:
A Slater insulator\cite{Slater1951} in the weak coupling regime in which the normal state is a non-magnetic metal. Below a critical temperature 
a gap in the single-particle opens up because of magnetic ordering and consequently the unit cell gets doubled and the Brillouin zone folds up. 
In this regime, the magnetic transition temperature increases with increasing Coulomb interaction strength.  
In the opposite strong coupling regime, the system is a Mott insulator\cite{Mott1949,Hubbard1963}. 
In this limit, the normal state is a gapped insulator due to strong Coulomb repulsion. 
Magnetic ordering sets in below a critical temperature $T_c\sim W^2/U$ related to the superexchange scale\cite{Anderson1950}. 
Interpolating between these two limits, the maximum $T_c$ is expected to occur in the crossover regime where $U$ and $W$ are comparable. 

The behavior of $T_c$ in a series of perovskite transition-metal oxides (TMOs) with the formal $t_{2g}^3$ electron configuration,  
including SrMnO$_3$, SrTcO$_3$ and NaOsO$_3$, lend support to the above picture.
$3d$ TMO SrMnO$_3$ has a relatively low N{\'e}el temperature $T_N=260$~K \cite{Takeda1974} for antiferromagnetic (AF) ordering 
but the insulating behavior persists above $T_N$, indicating that it is in the strong coupling limit.
On the other hand, $5d$ TMO NaOsO$_3$ with an AF transition $T_N=410$~K and a metal-insulator transition that is coincident \cite{Shi2009}
appears to be in the weak coupling limit. 
Finally, $4d$ TMO SrTcO$_3$ shows an extremely high AF $T_N\sim 1000$~K \cite{Rodriguez2011}, 
which indicates it is located in the crossover regime \cite{Mravlje2012}. 
It is also worth to note  that the crossover behavior associated with high magnetic transition temperatures is realized in double perovskite TMOs $A_2 BB'$O$_6$ \cite{Meetei2013}, 
where $A$ is an alkali or alkaline earth metal, and $B$ and $B'$ are two transition metals with different Coulomb interaction strengths. 

Given this backdrop, an important question that arises is {\it can there be systems that are incompatible with this classification?} 
And more broadly, what are the criteria for weak coupling Slater and strong coupling Mott regimes in multi-orbital systems?

One such systems is pyrochlore Cd$_2$Os$_2$O$_7$. 
As in NaOsO$_3$, the metal-insulator transition of  Cd$_2$Os$_2$O$_7$ accompanies a magnetic ordering \cite{Sleight1974,Mandrus2001}. 
However, unlike NaOsO$_3$ the unit cell of Cd$_2$Os$_2$O$_7$ contains four Os sites and therefore can become an insulator 
without changing the size of the unit cell. 
In fact, the observed all-in-all-out magnetic structure is compatible with the structural unit cell \cite{Yamaura2012,Shinaoka2012}. 
It turns out that the metal-insulator transition in Cd$_2$Os$_2$O$_7$ is a Lifshitz transition \cite{Lifshitz1960}
arising from a reconstruction of the Fermi surface, rather than a Slater transition. 

\begin{figure}
\begin{center}
\includegraphics[width=1\columnwidth, clip]{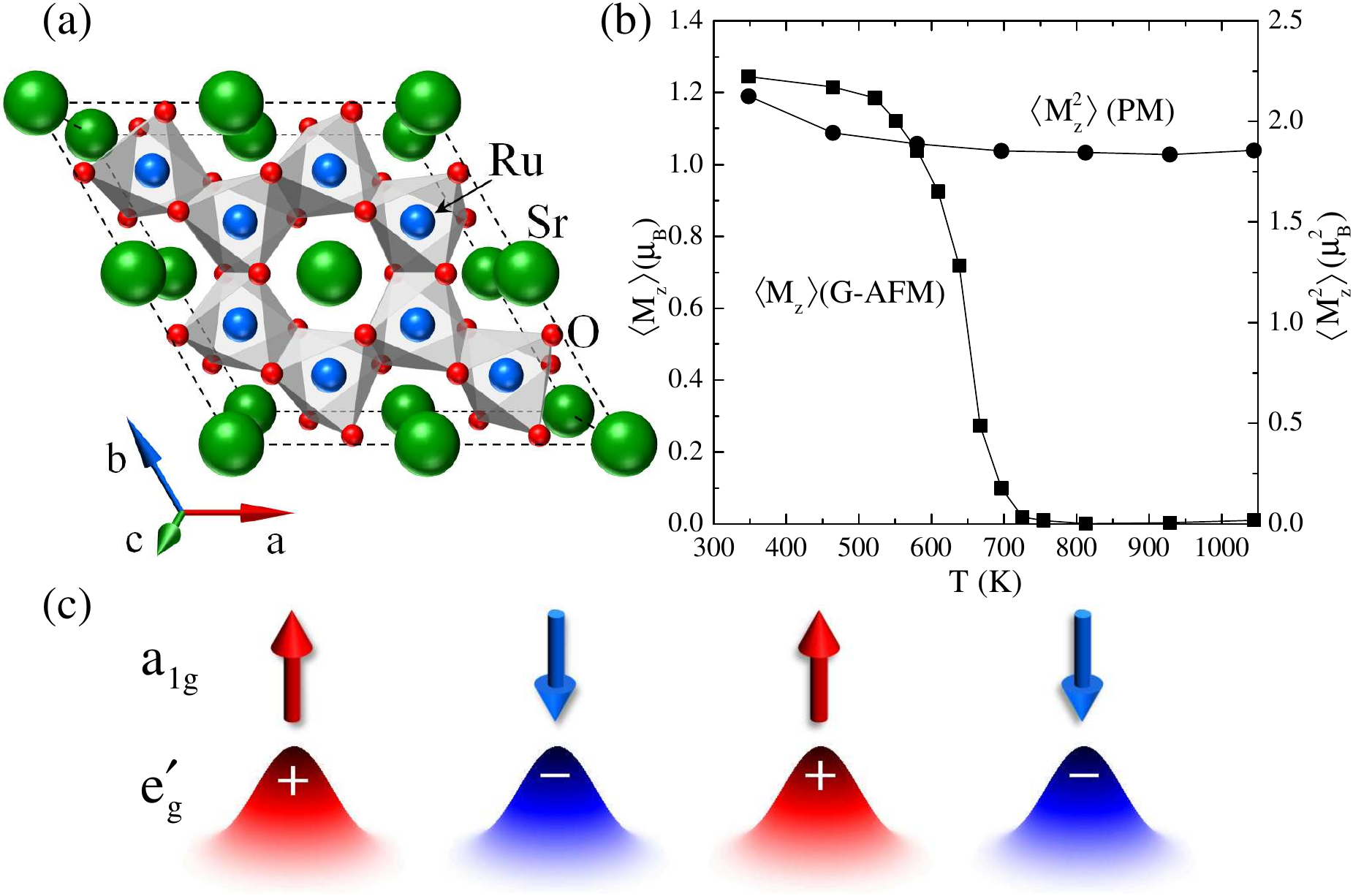}
\caption{(a) Top view of SrRu$_2$O$_6$. The cell is doubled along the $a$ and $b$ directions.
(b) Ordered $\langle M_z \rangle$ and equal-time spin-spin correlation $\langle M_z^2 \rangle$ 
as a function of temperature $T$. 
From the high temperature value, $\langle M_z^2 \rangle \sim 1.8 \mu_B^2$, 
the local moment is estimated to be $M \sim 2.3 \mu_B$.
(c) Schematic view of the localized vs. itinerant dichotomy between $a_{1g}$ electrons and $e'_g$ electrons. 
Correlation effects are stronger for $a_{1g}$ electrons resulting in localized moments, while 
$e'_g$ electrons maintain a rather itinerant character and show spin-density wave (SDW) like behavior when magnetically ordered.  
}
\label{fig:Fig1}
\end{center}
\end{figure}

In this paper, we focus on another compound, the hexagonal SrRu$_2$O$_6$ \cite{Hiley2014} [Fig. \ref{fig:Fig1} (a)]. 
Given a half filled $t_{2g}$ configuration for Ru sites from its formal valence $+5$, like Os$^{5+}$ in NaOsO$_3$ or Cd$_2$Os$_2$O$_7$, 
and also because Ru is a $4d$ element, one may expect that it would show behavior similar to SrTcO$_3$.  
Indeed, a fairly high $T_N \sim 565$~K is observed in SrRu$_2$O$_6$\cite{Tian2015}. 
However, the ordered moment $\sim 1.3 \mu_B$ is found to be much smaller than that expected for Ru$^{5+}$, $3 \mu_B$ \cite{Tian2015,Hiley2015}. 
To solve this puzzle, there have appeared a number of theoretical studies using density functional theory (DFT). 
Singh \cite{Singh2015} and Tian {\it et al.}\cite{Tian2015} argued that the strong hybridization between Ru and O ions is responsible for the reduction of the ordered moment. 
Streltsov and coworkers proposed the formation of molecular orbitals within a Ru hexagonal plane owing to such a strong hybridization \cite{Streltsov2015}.
They also performed the DFT calculations combined with the dynamical mean field theory \cite{Georges1996} (DFT+DMFT \cite{Anisimov1997a,Lichtenstein1998,Kotliar2006}). 
It was reported that the ordered moment on a Ru site becomes $3 \mu_B$ as anticipated for Ru$^{5+}$ and that the transition temperature becomes $\sim 2000$~K, 
which is very similar to a theoretical result for SrTcO$_3$ \cite{Mravlje2012} and in return supports the importance of the molecular orbital picture that is not considered in DMFT. 

In this paper, we investigate the electronic and magnetic properties of SrRu$_2$O$_6$ using DFT+DMFT. 
Our main finding is summarized in Fig. \ref{fig:Fig1} (b). 
In contrast to the previous report of Ref.~\cite{Streltsov2015}, 
we find that the ordered moment is indeed $\sim 1.2 \mu_B$, which is nearly identical to the DFT result $1.3 \mu_B$ \cite{Singh2015} 
and also consistent with the experimental value $1.4 \mu_B$ \cite{Tian2015,Hiley2015}. 
These observations might indicate that SrRu$_2$O$_6$ is in the weak coupling limit, similar to NaOsO$_3$, not in the crossover regime. 
However, the local moment estimated by the equal-time spin-spin correlation remains relatively large throughout the temperature range analyzed. 
We also find that the transition temperature estimated from DMFT is $T_N \sim 700~K$, reasonably close to the experimental  $T_N \sim 565~K$ \cite{Tian2015,Hiley2015}. 
Fluctuations not included in DMFT are expected to reduce the ordering temperature bringing it closer to the experimental estimate.
Even more significantly, we find that one of $t_{2g}$ states shows strong mass enhancement. 
Thus, our results indicate SrRu$_2$O$_6$ is in close proximity to an orbitally-selective Mott insulating state 
and localized and itinerant electrons coexist as schematically shown in Fig. \ref{fig:Fig1} (c). 

\section*{Results}

\begin{figure}
\begin{center}
\includegraphics[width=0.9\columnwidth, clip]{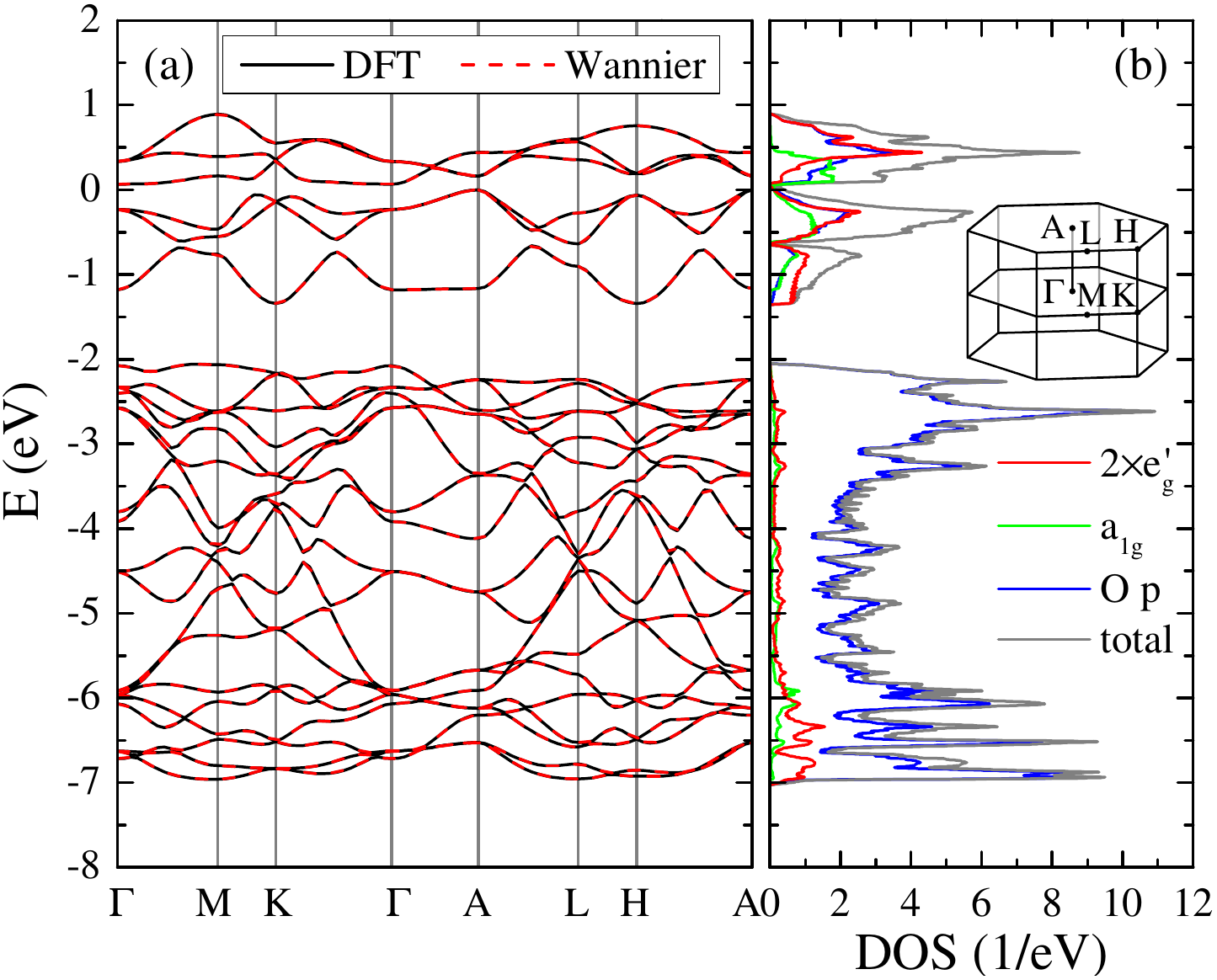}
\caption{DFT results on SrRu$_2$O$_6$. 
(a) Band dispersion (solid lines) compared with Wannier dispersions (broken lines).
(b) Total and orbital-resolved (Ru $e_g'$, $a_{1g}$ and O $p$) DOS. 
The valence-band maximum is set to $E=0$. 
DOS for $e_g'$ involves contributions from two degenerate bands.
The inset shows the first Brillouin zone for the hexagonal lattice. }
\label{fig:Fig2}
\end{center}
\end{figure}


\subsection*{DFT analysis}

We utilize DFT+DMFT techniques \cite{Anisimov1997a,Lichtenstein1998,Kotliar2006} 
to better account for correlation effects associated with the localized Ru $d$ electrons. 
We first perform the DFT calculations for the non-magnetic state with the Elk package \cite{elk}.
The experimental lattice parameters are considered with atomic configurations determined at room temperature \cite{Hiley2014}. 
It turns out that these calculations themselves provide important insights into the novel behavior of SrRu$_2$O$_6$. 
The spin-orbit coupling (SOC) $\lambda$ is neglected in our analyses because $\lambda \sim 200$~meV in ruthenium oxides \cite{Fatuzzo2015}
is smaller than our estimation for the Hund coupling parameters. 
We anticipate that the main effect of the SOC is to introduce the uniaxial spin anisotropy. 

Within the DFT, a small band gap $\sim 0.05$ eV exists at the Fermi level in a paramagnetic state \cite{Tian2015}, indicating that SrRu$_2$O$_6$ is a band insulator. 

We then construct a model Hamiltonian consisting of Ru $t_{2g}$ states and O $p$ states using the Wannier functions \cite{Marzari1997, Souza2001, Kunes2010, Mostofi2008} and transfer integrals between them on different sites extracted from the DFT band structure.
We estimate the interaction parameters from the constrained random phase approximation (cRPA) \cite{Aryasetiawan2004}.
Details of our model construction are provided in the Methods section. 
The interacting model thus obtained is solved by means of DMFT with the continuous-time quantum Monte-Carlo impurity solver \cite{Werner2006a,Werner2006b,Haule2007}.

It is instructive to start from the DFT results as summarized in Fig. \ref{fig:Fig2}. 
Fig. \ref{fig:Fig2} (a) shows the DFT dispersion relations near the Fermi level. 
The valence-band maximum is set to $E=0$. 
The first Brillouin zone for the hexagonal lattice is shown as an inset of Fig. \ref{fig:Fig2} (b). 
In this calculation, magnetic ordering is suppressed. 
RuO$_6$ octahedra are slightly compressed along the $c$ direction, and O-Ru-O angle is about 94.4$^\circ$ 
(two oxygen sites are on the same plane).  
Reflecting this distortion and the hexagonal symmetry, i.e., $a_{1g}$ and $e_g'$ are intrinsically inequivalent, 
the $a_{1g}$ level is about 0.3 eV higher than the $e_g'$ level in the Wannier basis. 

These bands are well separated from the other bands located below $-15$~eV or above $+3$~eV.  
As shown in Fig. \ref{fig:Fig2} (b), these bands primarily come from Ru $t_{2g}$ states
(split to twofold degenerate $e_g'$ states and non-degenerate $a_{1g}$ states) and O $p$ states, 
justifying our choice of the basis set of the model Hamiltonian. 
It is clearly seen that Ru $t_{2g}$ and O $p$ states are strongly hybridized in the energy window[$-7.0$:$+1.0$] eV, 
which is used to construct Wannier functions in this study. 
Because of this large energy window, Wannier orbitals with the $t_{2g}$ symmetry centered on a Ru site are well localized with the typical spatial spread 1.0--1.2 ~\AA. 
Interestingly, a substantial amount of O $p$ states are above the Fermi level, corresponding to 2.596 holes in the O $p$ states per unit cell or 0.433 holes per oxygen ion. 
This point will become crucial in the discussion below.  
It should also be mentioned that Ru $ a_{1g}$ states have the largest weight near the Fermi level 
(Note that Ru $e_g'$ states have the twofold degeneracy, therefore each $e_g'$ band has the smaller spectral weight than the $a_{1g}$ band near the Fermi level). 
This indicates that the effect of correlations is different between $e_g'$ and $a_{1g}$ states.

Figure \ref{fig:Fig2} (a) also shows the Wannier dispersion relations, which completely overlap with the DFT dispersion relations.
There are 48 bands including the spin degeneracy, 
i.e., 6 per Ru ($t_{2g}$) and 6 per O ($p_x, p_y, p_z$). 
A Sr does not contribute to these low-energy bands 
because Sr $5s (4p)$ level is so high (low) and its valence state is $+2$ by donating 2 electrons to these bands. 
Considering the nominal valence, Sr$^{2+}$Ru$_2^{5+}$O$_6^{2-}$, and the charge counting, 
3 per Ru$^{5+}$ and 6 per O$^{2-}$, these bands are filled by 42 electrons. 
In this study, the orbital occupancy is determined in the Wannier basis. This is also used to determine the valence state of a Ru ion and the magnetic moment. 
Within DFT with the Wannier basis, the filling is about 1.42 per $a_{1g}$ orbital and about 1.45 per $e_g'$ orbital, so these orbitals are nearly equally filled. 
Since the total $t_{2g}$ occupation 4.3 is larger than 3, the Ru valence state is deviated from the nominal value +5 to +3.7.

\subsection*{Double-counting correction}

\begin{figure}
\begin{center}
\includegraphics[width=0.7\columnwidth, clip]{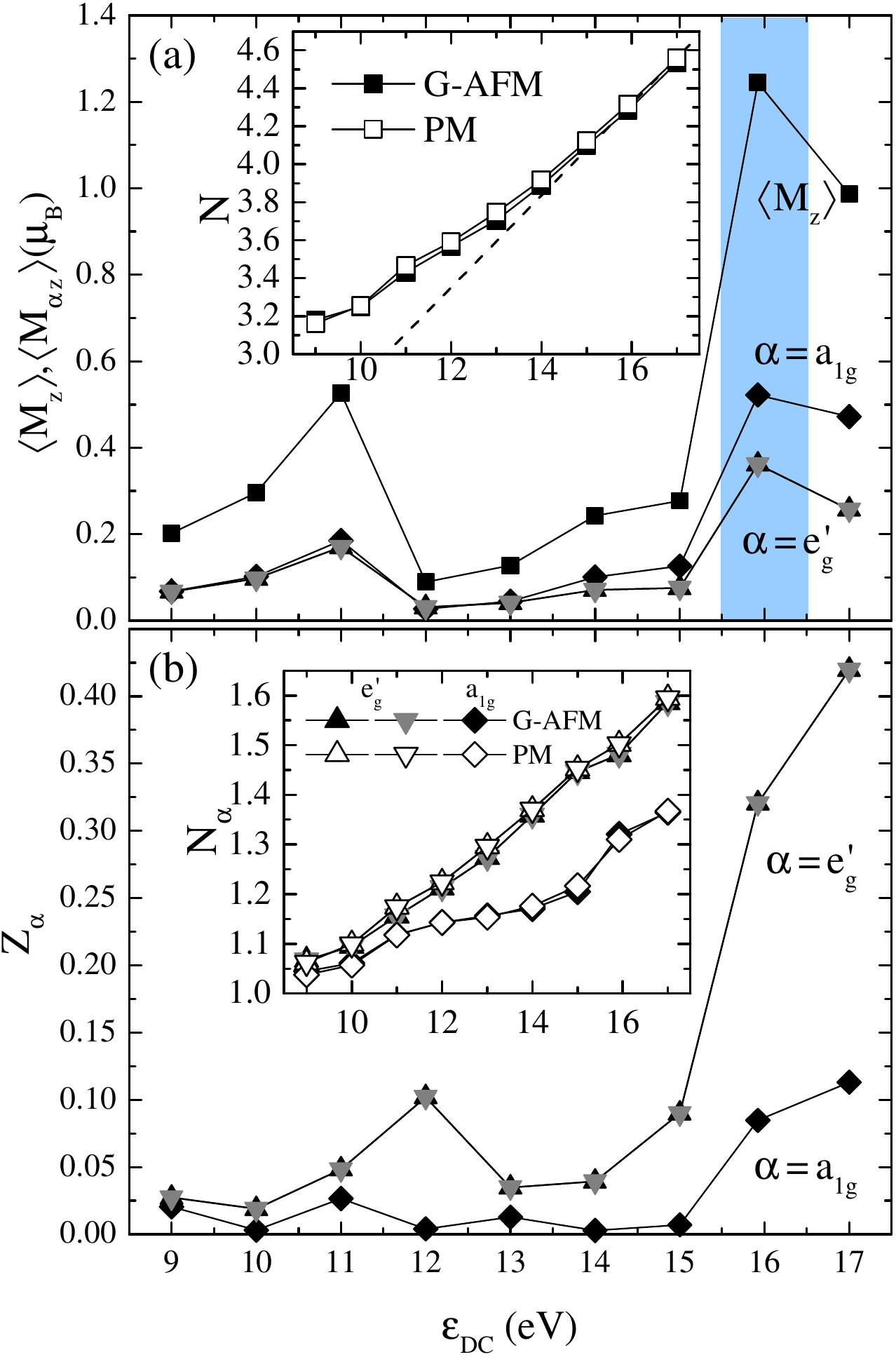}
\caption{(a) DMFT results of the total and the orbital-resolved ordered magnetic moments, 
$\langle M_z \rangle$ and $\langle M_{\alpha z} \rangle$, respectively, as a function of $\varepsilon_{DC}$. 
(b) Quasiparticle weight $Z_\alpha$ as a function of $\varepsilon_{DC}$. 
The insets of (a) and (b) respectively show the total $d$ electron density $N$ and 
the orbital-resolved density $N_\alpha$ as a function of $\varepsilon_{DC}$. 
}
\label{fig:Fig3}
\end{center}
\end{figure}


For methods using DFT supplemented by many-body calculations, 
it is critical that we first correct for the Hartree contribution arising from correlated orbitals that is already included in the DFT calculation; we call this the double-counting correction (DC). 
This is introduced as a uniform level shift of correlated orbitals $\varepsilon_{DC}$.
Two methods for accounting for such a DC correction have been investigated during the development of local (spin) density approximation [L(S)DA] $+U$ methods \cite{Czyzyk1994,Anisimov1997b}.
Since the combination of DFT+ DMFT techniques started to appear, a number of DC correction methods have been proposed
\cite{Lichtenstein2001,Amadon2008,Haule2010,Kunes2012,Haule2014,Haule2015}.
However, it remains unclear which of these methods is best suited for the current system. 
Here, instead of examining each DC method, we take a different approach: 
we treat $\varepsilon_{DC}$ as an adjustable parameter and, based on the experimental input, i.e., the ordered moment, 
determine the parameter range suitable for the material and then use it for the more detailed calculations of the quasi-particle weight.

The inset of Figure \ref{fig:Fig3} (a [b])shows the total (orbital-resolved) Ru $d$ electron densities $N(N_\alpha)$ as a function of $\varepsilon_{DC}$. 
As one can see, the total density $N$ remains larger than 3, the nominal value for Ru$^{5+}$, for the whole $\varepsilon_{DC}$ range examined.  
To realize $N=3$, i.e., the $d^3$ electron configuration, $\varepsilon_{DC}$ must be unrealistically small. 
$N$ and $N_\alpha$  are not sensitive to the magnetic state, 
compare paramagnetic (PM) solutions and G-type antiferromagnetic (G-AFM) solutions. 
In all $\varepsilon_{DC}$, $a_{1g}$ orbitals have electron density closer to 1 than $e_g'$ orbitals. 
From this, one could anticipate that $a_{1g}$ electrons contribute to magnetism more strongly than $e_g'$ electrons. 

%

In the inset of Fig. \ref{fig:Fig3} (a), we also plot the analytical line
\begin{equation}
\varepsilon_{DC} = U_{ave} \biggl(N-\frac{1}{2}\biggr) - \frac{1}{2} J_{ave} (N-1),
\label{eq:FLL}
\end{equation}
which is the so-called fully localized limit (FLL) \cite{Czyzyk1994}, with $U_{ave}$ and $J_{ave}$ being averaged Coulomb interactions and exchange interactions 
computed from matrix elements presented in the Method section, Eqs. (\ref{eq:U}) and (\ref{eq:J}), respectively. 
We notice that the analytic curve and the numerical data practically overlap at $\varepsilon_{DC} \agt 15$~eV. 
Thus, if one takes the FLL DC correction and imposes the condition that impurity occupation $N$ and the one in the DC correction Eq. (\ref{eq:FLL}) coincide, 
any point at $\varepsilon_{DC} \agt 15$~eV fulfills this condition. 
This means that the full convergence for this scheme is very difficult. 
Instead, we take the Ru $d$ occupancy from the DFT calculation, $N=4.3$, and use this value in Eq. (\ref{eq:FLL}) to obtain $\varepsilon_{DC}=15.92$~eV. 
This $\varepsilon_{DC}$ gives impurity occupation $N=4.28 (4.31)$ in the PM (G-AFM) solution at $T=348$~K.
It is reported that the FLL formula gives a slightly larger impurity occupancy than the ``nominal'' DC (in this case $N=3$), which is close to the ``exact'' DC \cite{Haule2015}. 
However, the true nominal occupancy in Ru $d$ states is not known yet. 
While it remains to be justified, our results using the FLL DC with $N=4.3$, as discussed below, 
suggest that our choice of $\varepsilon_{DC}$ provides a reasonable physical picture of SrRu$_2$O$_6$. 
It is very interesting to apply the improved DC correction to SrRu$_2$O$_6$, but this remains a future research. 

Given the vital information on the Ru $d$ occupation number $N$ vs. $\varepsilon_{DC}$ in the inset of Fig.~\ref{fig:Fig3} (a), 
we examine physical quantities as a function of $\varepsilon_{DC}$. 
Figure \ref{fig:Fig3} (a) shows the total ordered moment $\langle M_z \rangle$ and orbital-resolved ordered moment $\langle M_{\alpha z} \rangle$ 
for G-AFM as a function of $\varepsilon_{DC}$. 
Here, $\langle M_z \rangle = \sum_\alpha \langle M_{\alpha z} \rangle$ with $\langle M_{\alpha z} \rangle= N_{\alpha \uparrow}-N_{\alpha \downarrow}$. 
One notices that the ordered moment on the $a_{1g}$ orbital is larger compared to the $e_g'$ orbitals. 
This is because the $a_{1g}$ occupation is closer to one than the  $e_g'$ occupations [see the inset of Figure \ref{fig:Fig3} (b)] and, 
therefore, the spin polarization is easily induced. 
 
It is remarkable that the ordered moment varies drastically with $\varepsilon_{DC}$. 
This comes from the smaller occupation in $a_{1g}$ orbitals than $e_g'$ orbitals, i.e., $N_{a_{1g}} < N_{e_g'}$ [see Fig. \ref{fig:Fig3} (b) inset]. 
This is partly explained by DFT results; it is found that the bare $a_{1g}$ level is about $0.3$ eV higher than the $e_g'$ level, 
and  $N_{a_{1g}}=1.42 < N_{e_g'}=1.45$. 
However, the orbital polarization  $N_{e_g'}-N_{a_{1g}}$ is found to be significantly enhanced in our DMFT calculations; from 0.03 in DFT to at most 0.25. 
This is because the orbital polarization is enhanced by correlations \cite{Okamoto2004,Werner2007}. 
Furthermore, as shown later, spectral weight right below the Fermi level is dominated by $a_{1g}$ orbital [Fig. \ref{fig:Fig4} (a)]. 
Since reducing 
$\varepsilon_{DC}$ roughly corresponds to lowering $E_F$, 
$a_{1g}$ orbital changes the occupation number more sensitively than $e_g'$ orbitals, 
leading to the large suppression in the ordered moment with decreasing $\varepsilon_{DC}$. 

With $\varepsilon_{DC}=15.92$~eV (the FLL DC correction with $N=4.3$), we found the ordered moment $\langle M_z \rangle=\sim 1.2 \mu_B$. 
This value is consistent with the experimental observation $1.4 \mu_B$ \cite{Tian2015,Hiley2015}, giving us confidence in our choice of $\varepsilon_{DC}$. 
This is quite different from the nominal occupation of Ru$^{5+}$, $N=3$. 
Such a large difference of $N$ from its nominal value comes from the strong mixture between Ru $d$ states and O $p$ states 
in low-energy bands. 
It should be noted that in order to achieve $N=3$, one needs to adopt unphysically small $\varepsilon_{DC}$. 

The strong sensitivity of $\langle M_z \rangle$ on $\varepsilon_{DC}$ indicates the magnetism and the band structure are strongly coupled. 
This might suggest that SrRu$_2$O$_6$ is in the weak coupling regime despite relatively large values of the matrix elements of $\hat U$. 
But is this true? 
Thus, we now turn to the quasiparticle weight, which is the direct measure of the correlation strength.  
Figure \ref{fig:Fig3} (b) 
shows orbital dependent quasiparticle weight $Z_\alpha$ as a function of $\varepsilon_{DC}$. 
It is seen that $Z_\alpha$ for $\alpha = a_{1g}$ becomes extremely small at $\varepsilon_{DC} <10$ eV and $\varepsilon_{DC}\sim 14$ eV approaching 
integer fillings $N=3$ and 4, respectively,  
while $Z_\alpha$ for $\alpha = e_g'$ remains larger than 0.02. 
This indicates that the correlation effect is quite strong and the system is in the vicinity of orbital-selective Mott insulating regimes \cite{DeMedici2009}, 
in which $\alpha = a_{1g}$ electrons tend to be localized but $e_g'$ electrons maintain an itinerant character. 
This dichotomy between localized and itinerant characters will be discussed in detail later. 
%
%
We also note that the coexistence of localized and itinerant electrons would lead to the bad-metal behavior 
that has been suggested in an extended range of coupling and carrier densities in strongly-correlated multiorbital systems in the presence of Hund's coupling \cite{Medici2011,Georges2013}. 
 
In many magnetic materials, the ordered moment increases with increasing correlation strength. 
Thus, the quasiparticle weight in a PM solution and the ordered moment in a magnetic solution tend to have a negative correlation. 
As shown in Figs. \ref{fig:Fig3} (a) and (b), 
on the contrary, the ordered moments become large where the quasiparticle weights become large, i.e., they have a positive correlation. 
All these observations suggest that SrRu$_2$O$_6$ is in the strong coupling regime, but Mott physics and itinerant band physics coexist.


\subsection*{Temperature dependence}

Based on the detailed comparison with experiments, 
now we focus on $\varepsilon_{DC}=15.92$~eV and examine the temperature dependence of the electronic property of SrRu$_2$O$_6$. 

\begin{figure}
\begin{center}

\includegraphics[width=0.9\columnwidth, clip]{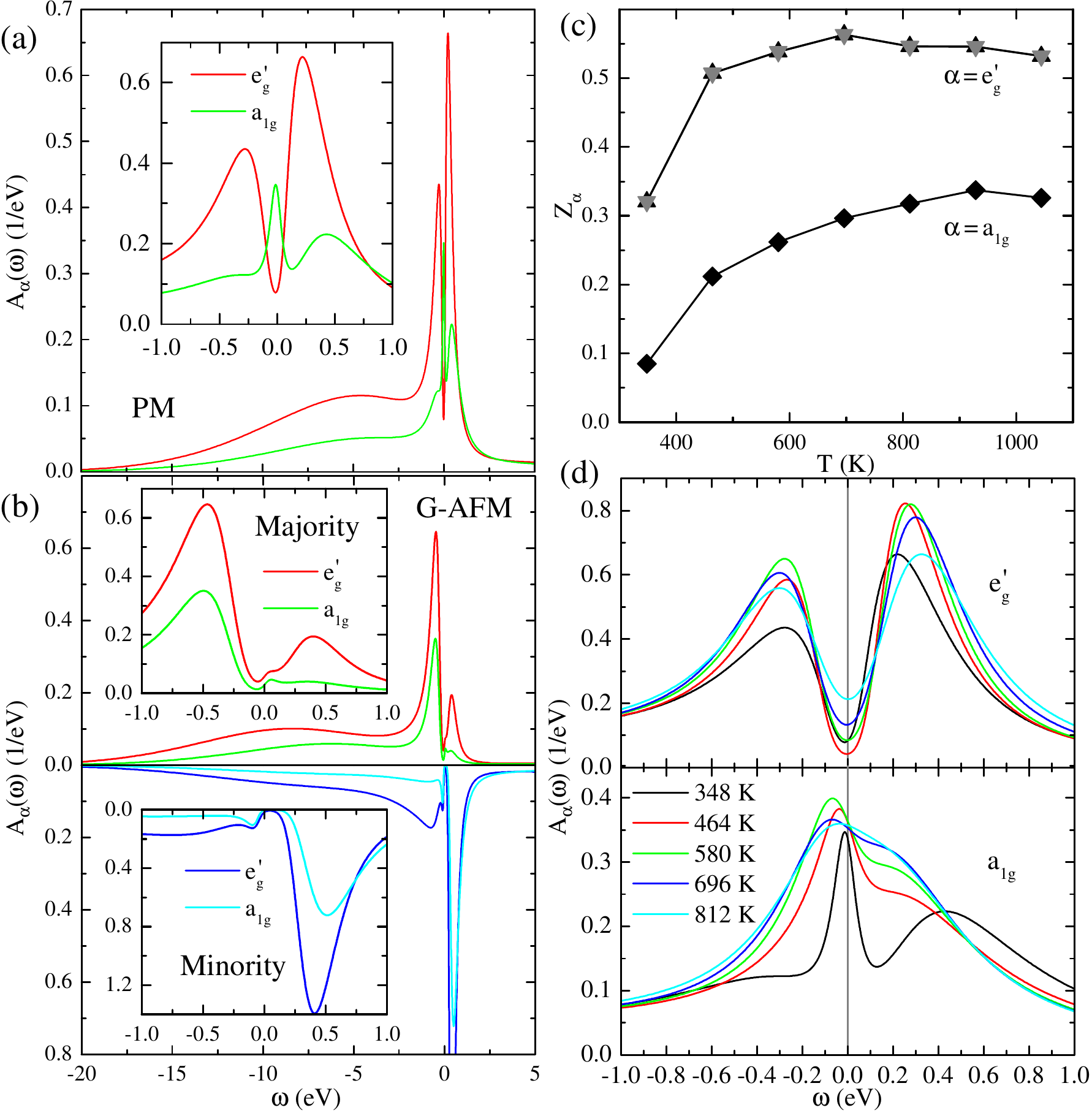}
\caption{DMFT results of the spectral function $A_\alpha (\omega)$ for a PM phase (a) and a G-AFM phase (b) at $T=348$~K. 
(c) Orbital-resolved quasiparticle wright $Z_\alpha$ as a function of $T$. 
The $a_{1g}$ band shows a stronger mass enhancement than the $e'_g$ band by a factor of 2-- 3. 
(d) Temperature dependent spectral function $A_\alpha (\omega)$.  
}
\label{fig:Fig4}
\end{center}
\end{figure}

We start from the low-temperature electronic property. 
Figure \ref{fig:Fig4} (a,b) 
show the Ru $d$ density of states computed by the maximum entropy analytic continuation for the impurity Green's functions on the Matsubara axis \cite{Jarrell1996}. 
$e_g'$ states have a clear gap in both PM and G-AFM solutions with the peak-to-peak distance $\sim 0.6$~eV and $\sim 0.9$~eV, respectively. 
For $a_{1g}$ states, the gap structure is rather vague especially in the PM solution. 
The minimum peak-to-peak distance is $\sim 0.2$~eV for both PM and G-AFM. 
These values are comparable to the experimental estimate on the activation energy $\sim 0.4$~eV \cite{Tian2015}.


Magnetic properties are summarized in Fig. \ref{fig:Fig1} (b). 
Local ordered moment $\langle M_z \rangle$ is computed for G-type AFM and plotted as a function of temperature. 
At a glance, one notices a close resemblance between the theoretical $\langle M_z \rangle$ vs. $T$ curve and experimental ones reported in Refs. \cite{Tian2015,Hiley2015}. 
Theoretical results for the ordered moment at low temperatures is $\langle M_z \rangle \sim 1.2 \mu_B$ and the N{\'e}el temperature $T_N \sim700$~K 
are reasonably close to the experimental results, $\langle M_z \rangle \sim 1.4 \mu_B$ and  $T_N\sim565$~K. 
Theoretical $T_N$ is overestimated by about $20$~\%, which could be attributed to the mean field nature of the current ``single-site'' DMFT 
and the uniaxial anisotropy existing in real material due to the finite spin-orbit coupling but absent in our calculations. 

In Slater-type systems in the weak coupling limit, local moments disappear when magnetic ordering disappears above $T_N$. 
On the other hand, in Mott insulators in the strong coupling limit, local moments remain unchanged above $T_N$. 
In the intermediate crossover regime, local moments decrease with increasing temperature but survive at high temperatures. 
We have also computed the equal-time spin-spin correlation in an impurity model $\langle M_z^2 \rangle$ 
by suppressing magnetic ordering and plotted in Fig. \ref{fig:Fig1} (b). 
It is remarkable that $\langle M_z^2 \rangle$ remains constant in the whole temperature range. 
From $\langle M_z^2 \rangle \approx 1.8~\mu_B^2$ and using the spin rotational symmetry, 
the size of the local moment is deduced as $M =\sqrt{3 \langle M_z^2 \rangle} \approx 2.3 \mu_B$, 
which is roughly 2 times larger than the size of the ordered moment at low temperatures.  
Clearly SrRu$_2$O$_6$ is not in the weak-coupling limit and most likely in the strong-coupling limit. 
Similar dichotomy between large local moments and small ordered magnetic moments  
was theoretically suggested for iron-based superconductors \cite{Hansmann2010,Toschi2012} and 
has indeed been experimentally reported \cite{Liu2012}. 

We also found other evidence of strong correlation effects from the temperature dependent quasiparticle weight. 
As shown in Fig. \ref{fig:Fig4} (c), 
$Z_{\alpha}$ in PM solutions are decreased with decreasing $T$. 
This behavior contradicts with what one expects for a good metal where the electron coherence is enhanced with decreasing temperature. 
It is also interesting to point out that the renormalization in $Z_{e_g'}$ is moderate for high temperatures. 
This is because $e_g'$ bands maintain the band-insulator-like character and have larger spectral weights away from $\omega=0$. 
On the other hand,  the renormalization in $Z_{a_1g}$ is significant. 
Thus, SrRu$_2$O$_6$ is in the vicinity of the correlation-induced orbital-selective Mott transition.

Our findings are remarkably different from the previous DFT+DMFT study on  SrRu$_2$O$_6$. 
Ref.~\cite{Streltsov2015} found a large Ru moment $\sim 3 \mu_B$ and an extremely high transition temperature $\sim 2000$~K. 
While sufficient details are not provided in Ref.~\cite{Streltsov2015}, we suspect from our calculations that the main difference arises from the inclusion of ligand O $p$ states
in our calculations that are crucial to describe the electronic properties of the late transition-metal oxides with deeper $d$ levels.

 %
 As a closely related system, SrTcO$_3$ was studied in Ref. \cite{Rodriguez2011}. 
The experimental ordered moment is found to be $2.13 (1.87) \mu_B$ at low (room) temperature. 
The ordered moment deduced by DFT is $1.3 \mu_B$. 
All these values are substantially smaller than $S= 3/2$ moment $3.87 \mu_B$. 
SrTcO$_3$ as well as SrMnO$_3$, both of which are $d^3$ systems, were examined using DFT+DMFT in Ref.~\cite{Mravlje2012}.
The ordered moment is found to be $2.5 \mu_B$ for SrTcO$_3$ and $3 \mu_B$ for SrMnO$_3$ at low temperatures.
These values are close to the experimental values. 
The local moment for the two materials are estimated from $\langle S_{zi}^2\rangle$ above $T_N$ 
to be $2.7 \mu_B$ for the Tc and $3.8 \mu_B$ for Mn compound. 
The fluctuating magnetic moment of Mn is close to the maximal  value $3.87 \mu_B$, 
while that for Tc is suppressed because of the more-itinerant nature of SrTcO$_3$. 
The reported dependence of ordered and local moments on the ``$B$'' site elements of perovskite Sr$B$O$_3$ 
is consistent with the Slater-Mott crossover for half-filled Hubbard models. 
Our results on SrRu$_2$O$_6$ are closer to SrTcO$_3$ than SrMnO$_3$. 
However, the reduction from the local moment to the ordered moment is much stronger $1.3/2.3 \sim 0.57$ for our SrRu$_2$O$_6$ 
as compared with $2.5/2.7 \sim 0.93$ for SrTcO$_3$. 
This clearly indicates the importance of the coexistence of localized and itinerant electrons in SrRu$_2$O$_6$. 
It would be also very instructive to see the difference between SrRu$_2$O$_6$ and other ruthenium compounds with the nominal Ru valence of $+5$. 
Double perovskite ruthenates Ba$_2$LaRuO$_6$ ($T_N=29.5$~K), Ca$_2$LaRuO$_6$ ($T_N=12$~K) \cite{Greatrex1979,Battle1983} and 
Sr$_2$YRuO$_6$ ($T_N=26$~K) \cite{Greatrex1979,Battle1984} are among such compounds. 
The high temperature fluctuating (low temperature ordered) moment of Ba$_2$LaRuO$_6$, Ca$_2$LaRuO$_6$ and Sr$_2$YRuO$_6$ was deduced as 4.00, 4.27 and $3.13 \mu_B$ 
(1.96, 1.92 and $1.85 \mu_B$), respectively. 
These values are factor 1.4--1.8 larger than the corresponding values obtained in this study for SrRu$_2$O$_6$. 
The difference between these double perovskite ruthenates and SrRu$_2$O$_6$ can be ascribed to the more localized nature of the double perovskite ruthenates, 
in which Ru sites are separated by rare earth ions, and the Ru valence  expected to be much closer to +5. 

\section*{Discussion}



In the previous sections, we discussed magnetic and electronic properties of SrRu$_2$O$_6$. 
Here we discuss some implications of our results to the transport property. 
Because of the large imaginary part of the self-energies, it remains challenging to identify the low-energy feature of the spectral function. 
Nevertheless, the temperature dependence of $A_\alpha (\omega)$ might give a hint. 
As shown in Fig. \ref{fig:Fig4} (d), 
$e_g'$ bands show the gapped feature and, therefore, do not contribute to the transport much. 
On the other hand, the spectral function for the $a_{1g}$ band is gradually diminished at $\omega \sim 0$ with decreasing temperature. 
Thus, we expect the electric resistivity $\rho (T)$ to increase with decreasing temperature in the whole temperature range, including $T \gg T_N$. 

In the weak-coupling picture, SrRu$_2$O$_6$ is a band insulator at both above and below $T_N$, 
the gap amplitude is fixed at $T>T_N$ and grows rapidly according to the magnetic order parameter at $T<T_N$. 
Therefore, the $\rho (T)$ curve is expected to have a kink at $T_N$. 
On the other hand, in our strong coupling picture, the spectral functions evolve dynamically even in the PM regime. 
Due to the large self-energy or scattering rate, details of low-energy features in $A(\omega)$ are smeared out. 
Thus, $\rho (T)$ is expected to change smoothly across $T_N$. 

To verify our results, it would be very important to experimentally examine the resistivity in the whole temperature range, from $T \ll T_N$ to $T \gg T_N$. 
Our preliminary results of the resistivity above room temperature appear to support this. 
Also, (angle resolved) photoemission spectroscopy measurements would provide direct evidence of the dynamical change in the spectral function, 
the evolution of the pseudogap feature to the full gap with decreasing temperature across $T_N$. 

The current work suggests that SrRu$_2$O$_6$ has to be considered as a different class of material 
than a series of $t_{2g}^3$ TMOs, including SrMnO$_3$, SrTcO$_3$ and NaOsO$_3$, 
where the low-energy behavior is well captured by multi(3)-orbital Hubbard models. 
To understand the unique behavior of SrRu$_2$O$_6$, it is essential to consider strong hybridization between Ru $d$ and O $p$ states. 
This could largely modify the $T$ vs. $U$ phase diagram by reducing both $T_N$ and the local moment $M_{loc}$, 
while the correlation strength in TM $d$ states remains strong. 
There might be a large number of materials that could be classified into a similar category as SrRu$_2$O$_6$. 

The main effect of the hybridization with O $p$ states is driving the electron configuration of Ru to $t_{2g}^4$ rather than $t_{2g}^3$. 
In this respect, it would be very interesting to look for systems with the similar electron configurations as SrRu$_2$O$_6$ but stronger SOC. 
As discussed in Ref.~\cite{Svoboda2017}, the $d^4$ electron configuration would nominally imply an atomic moment $J=0$, 
but, interatomic exchange can lead to the formation of a local moment and novel spin-orbital entangled magnetism. 



To summarize, we have investigated the novel electronic and magnetic properties of the hexagonal compound SrRu$_2$O$_6$ using DFT+DMFT. 
The small moment on Ru $\sim 1.2 \mu_B$ and the relatively high N{\'e}el temperature $\sim 700$~K are consistent with the experimental report, 
$\sim 1.4 \mu_B$ and $T_N \sim 565$~K. 
These seemingly contradictory characters of SrRu$_2$O$_6$ are caused by the strong hybridization between Ru $d$ states and ligand O $p$ states, 
which increases the Ru $d$ occupancy substantially from the nominal value 3. 
The strong Ru $d$-O $p$ hybridization does not imply a weak-coupling nature. 
In fact, SrRu$_2$O$_6$ is in the strong-coupling regime, on the verge of the orbital-selective Mott localization. 
In contrast to the DFT results, a band insulator when magnetic ordering is suppressed, 
we predict that the electron spectral functions evolve dynamically with temperature. 
In particular above N{\'e}el temperature, the spectral functions exhibit pseudogap features, which turn to full gap with decreasing temperature. 
To verify our predictions, further experimental studies are desirable, including temperature dependent photoemission spectroscopy measurements and resistivity measurements.

\section*{Methods}

DFT calculations were performed for the non-magnetic state with the Elk package \cite{elk} 
using the exchange-correlation functional proposed by Perdew {\it et al.}~\cite{PW92}.
We considered the experimental lattice parameters and atomic configurations determined at room temperature \cite{Hiley2014}. 

Using the obtained band structure, 
we constructed the Wannier functions \cite{Marzari1997, Souza2001, Kunes2010, Mostofi2008}
containing the Ru $t_{2g}$ and O $p$ orbitals and calculated the transfer integrals among them. 
The energy window was set as [$-7.0$:$+1.0$] eV. 
Further, we evaluated the interaction parameters, the Coulomb repulsion $\hat{U}$ and the exchange coupling $\hat{J}$,
by the constrained random phase approximation (cRPA) \cite{Aryasetiawan2004}. 
In the calculation of the partially screened Coulomb interaction, we took 120 unoccupied bands and used a $4\times 4\times 4$ grid. 
The double Fourier transform of the constrained susceptibility was done with the cutoff of 5 (1/a.u.).
We neglected the SOC since it is small compared with the Hund coupling constants.
These calculations were performed using a density response code~\cite{Kozhevnikov2010} recently 
developed for the Elk branch of the original \textsc{exciting fp-lapw} code~\cite{elk}.

The obtained effective model consists of two parts as $H = H_{band} + \sum_{i \in {\rm Ru}} H_{int,i}$. 
The band part $H_{band}$ is given by 
\begin{eqnarray}
H_{band} = \sum_{\vec k,\sigma}\Bigl[\hat d_{\vec k \sigma}^\dag, \hat p_{\vec k \sigma}^\dag \Bigr] 
\Biggl[
\begin{matrix}
\hat \varepsilon^{dd}_{\vec k}-\varepsilon_{DC} \hat 1 & \hat \varepsilon^{dp}_{\vec k}\\
\hat \varepsilon^{pd}_{\vec k}& \hat \varepsilon^{pp}_{\vec k}
\end{matrix}
\Biggr]
\Biggl[
\begin{matrix}
\hat d_{\vec k \sigma}\\ \vec p_{\vec k \sigma}
\end{matrix}
\Biggr].
\end{eqnarray}
Here, $\hat d_{\vec k \sigma}$ ($\hat p_{\vec k \sigma}$) is a vector consisting of annihilation operators of Ru $t_{2g}$ (O $p$) electrons with spin $\sigma$ in a momentum $\vec k$ space. 
The dispersion $\hat \varepsilon_{\vec k}^{\alpha \beta}$ consists of Wannier parameters, with $\alpha$ and $\beta$ running through Ru $t_{2g}$ and O $p$ states. 
$\varepsilon_{DC}$ for interacting Ru $t_{2g}$ states is the DC correction. 

The interaction part $H_{int}$ is given by  
\begin{eqnarray}
H_{int} \!\! &=& \!\! 
\sum_a U_{aa} d_{a \uparrow}^\dag d_{a \uparrow} d_{a \downarrow}^\dag d_{a \downarrow}  
+ \sum_{a \ne b} U_{ab} d_{a \uparrow}^\dag d_{a \uparrow} d_{b \downarrow}^\dag d_{b \downarrow} \nonumber \\
&& +\sum_{a > b, \sigma} \bigl( U_{ab}-J_{ab} \bigr) d_{a \sigma}^\dag d_{a \sigma} d_{b \sigma}^\dag d_{b \sigma} \nonumber \\
&& +\sum_{a \ne b} J_{ab} \Bigl( d_{a \uparrow}^\dag d_{b \uparrow} d_{b \downarrow}^\dag d_{a \downarrow} 
+d_{a \uparrow}^\dag d_{b \uparrow} d_{a \downarrow}^\dag d_{b \downarrow}\Bigr). 
\end{eqnarray}
Here, the index $i$ for Ru sites is suppressed for simplicity. 
The matrix elements of $\hat U$ and $\hat J$, evaluated by using the cRPA, are given by 
\begin{eqnarray}
\hat U = 
\left[ \begin{matrix}
5.236 & 4.260 & 4.196 \\
4.260 & 5.236 & 4.196 \\
4.196 & 4.196 & 5.482
\end{matrix}
\right]
\label{eq:U}
\end{eqnarray}
and
\begin{eqnarray}
\hat J = 
\left[ \begin{matrix}
   & 0.488 & 0.567 \\
0.488 &   & 0.567 \\
0.567 & 0.567 & 
\end{matrix}
\right]
\label{eq:J}
\end{eqnarray}
in units of eV 
with the basis of $\{e'_{g1}, e'_{g2}, a_{1g} \}$. 

Recently, we become aware of a preprint by Hariki {\it et al}. (Ref.~\cite{Hariki}). 
They performed similar DFT+DMFT calculations with different interaction parameters and a DC scheme and obtained similar conclusion as Ref. \cite{Streltsov2015}.

\section*{Acknowledgments}
The research by S.O. and J.Y. is supported by 
the U.S. Department of Energy,  Office of Science, Basic Energy Sciences, Materials Sciences and Engineering Division.
This work was supported by JSPS KAKENHI Grants No. 15K17724 (M.O.) and 15H05883 (R.A.).
N.T. acknowledges funding from DOE BES Grant DE-FG02-07ER46423.
This research was initiated at the Kavli Institute for Theoretical Physics (KITP), the University of California, Santa Barbara, 
where three of the authors (S.O, R.A. and N.T.) attended the program 
``New Phases and Emergent Phenomena in Correlated Materials with Strong Spin-Orbit Coupling.''
S.O., R.A. and N.T. thank the KITP, which is supported in
part by the National Science Foundation under Grant No. NSF PHY11-25915, for hospitality. 

\section*{Author contributions}
S.O. and N.T conceived the project, S.O. performed the DMFT calculations, M.O. and R.A. performed the DFT calculations, 
S.O. and N.T wrote the manuscript., and  J.Y. provided experimental input. 
All the authors discussed the results.


\section*{Additional Information}
{\bf Competing financial interests:} The authors declare no competing financial interests.

\end{document}